\documentstyle[12pt]{article}
\newcommand{\eq}{\begin{equation}}
\newcommand{\en}{\end{equation}}
\newcommand{\eqn}{\begin{eqnarray}}
\newcommand{\enn}{\end{eqnarray}}

\newcommand{\nn}{\nonumber}

\begin{document}
\begin{titlepage}
\begin{flushright}
IASSNS-HEP-92/86 \\
Dec. 1992
\end{flushright}
\begin{center}
{\LARGE
Generalized Conformal and  \\
Superconformal Group Actions \\
and Jordan Algebras \\ }

\vspace{1cm}
{\large Murat G\"{u}naydin$^*$} \\
School of Natural Sciences \\
Institute for Advanced Study \\
Princeton, NJ 08540 \\
and \\
Penn State University$^{\dagger}$ \\
Physics Department \\
University Park, PA 16802 \\
{\bf Abstract}
\end{center}
\small
We study the ``conformal groups'' of Jordan algebras along the lines suggested
by Kantor. They provide a natural generalization of the concept of conformal
transformations that leave 2-angles invariant to spaces where ``$p$-angles''
($p \geq 2$) can
be defined. We give an oscillator realization of the generalized
conformal groups of
Jordan algebras and Jordan triple systems. A complete list of the
generalized conformal algebras of simple Jordan algebras and hermitian Jordan
triple systems is given.
 These results are then extended
to Jordan superalgebras and super Jordan triple systems. By going to a
coordinate representation of the (super)oscillators one then obtains the
differential
operators representing the action of these generalized (super) conformal groups
on the corresponding (super) spaces. The superconformal algebras of the Jordan
superalgebras in Kac's classification is also presented.\\

\normalsize
\footnoterule
\noindent
{\footnotesize ($^*$) Work supported in part by the
National Science Foundation Grant PHY-9108286. \\
$(^{\dagger})$ Permanent Address. \\
e-mail: Murat@psuphys1.psu.edu or GXT@PSUVM.bitnet }
\end{titlepage}
\setcounter{footnote}{0}
\section{Introduction}
\setcounter{equation}{0}
Conformal symmetry plays an important role in the formulation
and understanding of many
physical theories. For example, the massless gauge theories
in four dimensions are invariant
under the fifteen parameter conformal group $SO(4,2)$. The known string
theories are all invariant under the infinite conformal group in two
dimensions which can be identified with the reparametrization
invariance of the string world-sheet. The two dimensional physical systems
are known to exhibit conformal symmetry at their critical points.

The conformal invariance is normally defined as the invariance of a quadratic
form or a metric up to an overall scale factor which is a function of the
local coordinates. This implies , in particular, the local invariance of
angles defined by the metric or the quadratic form.
It would be of physical interest to know if there exist generalizations of
conformal invariance
to spaces that are naturally endowed with higher order forms.
For example, the $p$-brane theories are naturally endowed with a volume form
which is of order $p$. Such a generalization was suggested by Kantor
in his study of the invariance groups of ``$p$-angles'' that can be defined
over spaces with a $p$-form (not to be confused with a differential $p$-form)
\cite{IK}.
 He studied , in particular, the invariance groups
of the ``$p$-angles'' defined by Jordan algebras with a generic norm of degree
$p$. We shall refer to these groups generically as generalized conformal
groups. In this paper we shall give a simple oscillator realization of the
generalized conformal algebras of Jordan algebras. In this realization the
Jordan triple
product plays a crucial role. The fact that the Jordan triple product rather
than the binary Jordan product is essential in our formulation allows us to
extend our results to Jordan triple systems. We give a complete list of simple
Jordan algebras and hermitian Jordan triple systems and their generalized
conformal
algebras. We then extend our results to study the superconformal algebras of
Jordan superalgebras and super Jordan triple systems. The list of simple
Jordan superalgebras as classified by Kac \cite{VK} and their generalized
conformal superalgebras are also given.
By going to the coordinate representation of the oscillators one obtains
a differential operator realization of the action of the generalized
conformal and superconformal algebras on the corresponding spaces and
superspaces.

\section{Linear Fractional Groups of Jordan Algebras as Generalized
Conformal Groups}
\setcounter{equation}{0}
The conformal transformations $T$ on a Riemannian manifold with the metric
$g_{\mu\nu}$ are defined such that under their action the metric transforms
as
\eq
T :\hspace{1.0cm}  g_{\mu\nu} \longrightarrow  \phi g_{\mu\nu} \label{eq:c1}
\en
where $\phi$ is a scalar function of the coordinates. On a d-dimensional
Euclidean space ($d>2$) with a non-degenerate
positive definite quadratic form $(x,x)$
the conformal transformations leave invariant the following cross-ratio
associated with any set of four vectors $x,y,z,w$:
\eq
\frac{(x-z,x-z)}{(x-w,x-w)} \frac{(y-w,y-w)}{(y-z,y-z)} \label{eq:c2}
\en
as well as the quantity
\eq
\frac{(x,y)^2}{(x,x)(y,y)} \label{eq:c3}
\en
which is the cosine square of the angle between the vectors
$x$ and $y$. Over a Euclidean
space one may use (\ref{eq:c1}), (\ref{eq:c2}) or (\ref{eq:c3})
 interchangeably to define
conformal transformations. Using the condition (\ref{eq:c2}) or
(\ref{eq:c3})  allows for an interesting generalization of conformal
transformations as was suggested by Kantor
\cite{IK} , which we shall briefly review below.

Kantor considers an n-dimensional vector space $V$ endowed
with a non-degenerate
form of degree $p$
\eq
N(x)\equiv N(x,x,..,x)
\en
 To every ordered set of four vectors $x,y,z$ and $w$ in $V$
one associates a cross-ratio
\eq
\frac{N(x-z)}{N(y-z)} \frac{N(y-w)}{N(x-w)} \label{eq:c4}
\en
and for each set of $p$ straight lines ($p$-angle) with direction vectors
$x_{1},...,x_{p}$ one defines the quantity
\eq
\frac{N(x_{1},..x_{p})^p}{N(x_{1})N(x_{2})\cdots N(x_{p})} \label{eq:c5}
\en
which is called the measure of the $p$-angle.
Let us denote the invariance groups of the cross-ratio (\ref{eq:c4}) and
the measure of the $p$-angle (\ref{eq:c5}) $G$ and $\tilde{G}$, respectively.
It can be shown that if $\tilde{G}$ is finite dimensional then it is
isomorphic to $G$ \cite{IK}. Some of the most interesting realizations of the
above generalization of conformal transformations  are provided by Jordan
algebras with a norm form. If $J$ is a semi-simple Jordan algebra with a
generic norm form as defined by Jacobson \cite{NJ} then considering $J$ as
 a vector space  one can study its generalized conformal transformation
groups using the definitions (\ref{eq:c4}) or (\ref{eq:c5}).
The corresponding groups $G$ and $\tilde{G}$ coincide if and only if
$J$ contains no one dimensional ideals in the complex case and no one or
two dimensional ideals in the real case. The action of $G$ on $J$ can
be written as a ``linear fractional transformation'' of $J$. The linear
fractional transformation groups of Jordan algebras were studied by Koecher
\cite{MK} and  Kantor \cite{IK}.
 The linear fractional transformation groups of Jordan superalgebras
were studied in \cite{MG75,MG80,MG90,MG91}.
\section{Conformal Algebras of Jordan Algebras and Jordan Triple Systems}
\setcounter{equation}{0}
The  reduced
 structure group $H$ of a Jordan algebra $J$
 is defined as the invariance group
of its norm form $N(J)$. By adjoining to it the constant scale transformations
one gets the full structure group of $J$. The Lie algebra $g$ of the
conformal group of $J$ can be given a three-graded structure with respect to
the Lie algebra $g^0$ of its structure group
\eq
g = g^{-1} \oplus g^0 \oplus g^{+1}
\en
  where
\eq
g^0 = h \oplus E
\en
with $h$ denoting the Lie algebra of $H$ and $E$  the generator
of the constant scale transformations.
The Tits-Kantor-Koecher (TKK)\cite{TKK}
construction of the Lie algebra $g$
 establishes a one-to-one mapping
between the grade $+1$ subspace of $g$ and the corresponding Jordan algebra
$J$:
\begin{equation}
U_{a} \in g^{+1}  \Longleftrightarrow  a \in J
\end{equation}
Every such Lie algebra $g$ admits
 a conjugation (involutive automorphism) $\dagger$
under which the elements of the grade $+1$ subspace get mapped into the
elements of the grade $-1$ subspace.
\begin{equation}
U^{a} = U_{a}^{\dagger}  \in g^{-1}
\end{equation}
One then defines
\begin{equation}
\begin{array}{l}
[U_{a},U^{b}] = S_{a}^{b} \\
\  \\
{[}S_{a}^{b}, U_{c}{]} = U_{(abc)}
\end{array}
\end{equation}
where $S_{a}^{b} \in g^{0}$ and $(abc)$ is the Jordan
triple product
\eq
(abc) = a \cdot (b\cdot c) + (a\cdot b)\cdot c - b \cdot (a \cdot c)
\label{eq:JTP}
\en
with $\cdot$ denoting the commutative Jordan product.
 Under conjugation $\dagger$ one finds
\begin{equation}
\begin{array}{l}
(S_{a}^{b})^{\dagger} = S_{b}^{a} \\
\  \\
{[}S_{a}^{b},U^{c}{]} = - U^{(bac)}
\end{array}
\end{equation}
The Jacobi identities in $g$ are satisfied if and only if the triple
product $(abc)$ satisfies the identities
\begin{equation}
\begin{array}{l}
(abc) = (cba)  \\   \label{eq:JTS}
\  \\
(ab(cdx))-(cd(abx))-(a(dcb)x)+((cda)bx) = 0
\end{array}
\end{equation}
These identities follow from the defining identities of
a Jordan algebra:
\eq
a \cdot b = b\cdot a
\en
\eq
a \cdot (b \cdot a^2) = (a\cdot b) \cdot a^2
\en
The elements $S_{a}^{b}$ of  the structure algebra $g^0$ of $J$ satisfy :
\begin{equation}
[S_{a}^{b},S_{c}^{d}] = S_{(abc)}^{d}-S_{c}^{(bad)}=S_{a}^{(dcb)}
-S_{(cda)}^{b}
\end{equation}
  Denoting as $J_{n}^{A}$ and as $\Gamma(d)$
 the Jordan algebra of $n\times
n$   Hermitian matrices over the division algebra $A$,
and the Jordan algebra of Dirac gamma matrices
in $d$ Euclidean dimensions, respectively, one finds
the following conformal groups ($G$) and
reduced structure groups ($H$) of simple Jordan algebras:

\begin{displaymath}
\begin{array}{|c|c|c|}
\hline
&& \\
J & H & G \\
\hline
&& \\
J_{n}^{R} & SL(n,{\bf R}) & Sp(2n,{\bf R})\\
&& \\
J_{n}^{C}  & SL(n,{\bf C}) & SU(n,n) \\
&& \\
J_{n}^{H}  & SU^{*}(2n) & SO^{*}(4n) \\
&& \\
J_{3}^{O}  & E_{6(-26)} & E_{7(-25)} \\
&& \\
\Gamma(d) & SO(d,1) & SO(d+1,2) \\
\hline
\end{array}
\end{displaymath}

The symbols $ {\bf R} $,
 ${\bf C}$, ${\bf H}$, ${\bf O}$
 represent the four division  algebras.
We should also note that by taking different real forms of the Jordan
algebras one obtains different real forms of the conformal and reduced
 structure groups.

 In the TKK construction only the triple product $(abc)$ enters and
the identities (\ref{eq:JTS})
 turn out to be the defining identities of a Jordan
triple system (JTS). Therefore, the TKK construction extends trivially to
Jordan triple systems. Of particular interest are the hermitian JTS's for
which the triple product $(abc)$ is linear in the first and the last
arguments and anti-linear in the second argument.
 There exist  four infinite families
of hermitian JTS's and two exceptional ones
\cite{OL}. They are:

{\em Type $I_{P,Q}$ } generated by $P\times Q$ complex matrices
$M_{P,Q}({\bf C})$ with the triple product
\begin{equation}
(abc) = a b^{\dagger} c + c b^{\dagger} a   \label{eq:J1}
\end{equation}
where $\dagger$ represents the usual hermitian conjugation.

{\em Type $II_{N}$}
 generated by complex anti-symmetric
  $N \times N$ matrices $A_{N}({\bf C})$ with the ternary product
\ref{eq:J1}.

{\em Type $III_{N}$} generated by complex $N \times N$
 symmetric matrices $S_{N}({\bf C})$ with the product
\ref{eq:J1}.

{\em Type $IV_{N}$} generated by Dirac gamma matrices
$\Gamma_{N}({\bf C})$ in N dimensions
 with complex coefficients and the Jordan triple product
\eq
(abc)= a\cdot (\bar{b} \cdot c) + c \cdot (\bar{b} \cdot a)
-(a \cdot c) \cdot \bar{b} \label{eq:HJTP}
\en
where the bar $ -$ denotes complex conjugation.

{\em Type V } generated by $1 \times 2 $ complex octonionic
matrices $M_{1,2}({\bf O_{C}})$ with the triple product
\begin{equation}
(abc) = \{ (a \bar{b}^{\dagger}) c + (\bar{b} a^{\dagger}) c
- \bar{b} (a^{\dagger} c) \} + \{ a \leftrightarrow c \}
\end{equation}
where $\dagger$ denotes octonion conjugation times transposition.

{\em Type VI} generated by the exceptional Jordan algebra
of $3\times 3$ hermitian octonionic matrices $J_{3}^{{\bf O}}(
{\bf C})$ taken over the complex numbers with the triple product
\ref{eq:HJTP}.

Below we tabulate the simple hermitian JTS's and their conformal ($G$) and
reduced structure groups ($H$):
\begin{displaymath}
\begin{array}{|c|c|c|} \hline
HJTS & G & H  \\ \hline
\ & \ & \ \\
M_{P,Q}({\bf C}) & SU(P,Q) & SL(P,{\bf R})\times SL(Q,{\bf R})  \\
\ & \ & \  \\
A_{N}({\bf C}) & SO(2N)^* & SU^*(N)   \\
\ & \ & \  \\
S_{N}({\bf C}) & Sp(2N,{\bf R}) & SL(N,{\bf R})  \\
\ & \ & \\
\Gamma_{N}({\bf C}) & SO(N+1,2) & SO(N,1)    \\
\  & \ &  \\
M_{1,2}({\bf O_{C}}) & E_{6(-14)} & SO(8,2)  \\
\ & \ & \\
J_{3}^{{\bf O}}({\bf C}) & E_{7(-25)} & E_{6(-26)}       \\  \hline
\end{array}
\end{displaymath}

\section{Oscillator Realization of the Generalized Conformal Groups }
\setcounter{equation}{0}
In the TKK construction of the conformal algebras of Jordan algebras
and Jordan triple systems the commutation relations are  expressed
in terms of the triple product $(abc)$. Let us choose a basis $e_a$
for the Jordan algebra or the JTS and introduce the structure constants
$\Sigma_{ab}^{cd}$
for the Jordan triple product
\eqn
(e_a e_b e_c)& =& \Sigma_{ac}^{bd} e_d \\
&& \nn \\
a,b,..& =& 1,2,...D \nn
\enn
Using these structure constants one can give oscillator realizations
of the generalized conformal algebras . Consider now a set of $D$
bosonic oscillators $A_a,A^b (a,b,... = 1,2,...,D)$
 that satisfy the canonical commutation
relations:
\eqn
{[ A_a, A^b ]}& =& \delta_a^b \nn \\
&& \nn \\
{[ A_a, A_b ]}&=&0 \\
&& \nn \\
{[ A^a, A^b ]} &=&0  \nn
\enn
The bilinears
\eq
S_a^b  = - \Sigma_{ac}^{bd} A^c A_d
\en
generate the structure algebra  of the corresponding Jordan algebra or
the Jordan triple system under commutation
\eq
{[ S_a^b , S_c^d ]} = - \Sigma_{ca}^{de} S_e^b + \Sigma_{ce}^{db} S_a^e
\en
If we further let
\eq
U_a = - A_a
\en
and define
\eq
U^a = \frac{1}{2} \Sigma_{cd}^{ae} A^c A^d A_e
\en
we find that they close into the generators of the structure algebra
under commutation
\eq
{[ U_a , U^b ]} = S_a^b
\en
Furthermore they satisfy
\eqn
{[ S_a^b , U_c ]} &=& \Sigma_{ac}^{be} U_e \\
&& \nn \\
{[ U^a , U^b ]}& =& {[ U_a, U_b ]} =0
\enn
In proving some of these commutation relations we used the identity
\eq
\Sigma_{ce}^{df} \Sigma_{af}^{bg} - \Sigma_{ae}^{bf} \Sigma_{cf}^{dg}
= \Sigma_{cf}^{db} \Sigma_{ae}^{fg} - \Sigma_{ca}^{df} \Sigma_{fe}^{bg}
\en
and the symmetry of the structure constants $\Sigma_{ab}^{cd} =
 \Sigma_{ba}^{cd}$
, which
follow from the defining identities (\ref{eq:JTS}) of JTS's.
Thus the operators $U_a , U^a $ and $S_a^b$ generate the conformal algebra
of the corresponding Jordan algebra or the JTS. To see how this realization
is related to the action of the structure algebra on the JTS (or the Jordan
 algebra) expand the elements $x \in J$ in the basis $(e_a)$:
\eq
x = x^a e_a   \hspace{1.0cm} x \in{J}
\en
Then the action of the conformal algebra on $J$ is equivalent to the
action of the differential operators on the ``coordinates'' $x^a$
 obtained by realizing the operators
$A_a$ and $A^a$ as
\eqn
A_a &=& \frac{\partial}{\partial x^a} \\
&& \nn \\
A^a &=& x^a
\enn
This leads to the differential operator realization
\eqn
U_a& =& - \frac{\partial}{\partial x^a} \\
&& \nn \\
U^a & = & \frac{1}{2} \Sigma_{cd}^{ab} x^c x^d \frac{\partial}{\partial x^b} \\
&& \nn \\
S_a^b &=& - \Sigma_{ac}^{bd} x^c \frac{\partial}{\partial x^d}
\enn
Thus we can interpret the action of the generalized conformal groups on
Jordan algebras or JTS's in the usual way \cite{MG90,MG91} i.e.
the $U_a$ and $U^a$ are the generators of translations and ``special conformal
transformations'', respectively. The $S_a^b$ are the generators of
``Lorentz transformations'' and dilatations \cite{MG75,MG90,MG91}.

\section{Oscillator Realization of the Conformal Superalgebras of
Jordan Superalgebras}
\setcounter{equation}{0}
Jordan superalgebras were defined and classified by Kac \cite{VK} using methods
developed for Jordan algebras by Kantor \cite{IK2}. An infinite family of
Jordan superalgebras was missed in the classification of Kac and was discovered
by Kantor \cite{IK3}. A Jordan superalgebra $J$ is a
$Z_2$ graded algebra
\eq
J=J^0 \oplus J^1
\en
 with a supercommutative product
\eqn
A \cdot B &=& (-1)^{d_A d_B} B \cdot A \\
&& \nn \\
d_A , d_B , ...& =& 0,1  \nn
\enn
that satisfies the super Jordan identity:
\eq
(-1)^{d_A d_C} {[} L_{A\cdot B} , L_C \} +
(-1)^{d_B d_A} {[} L_{B \cdot C}, L_D \} +
(-1)^{d_C d_B} {[} L_{C\cdot A}, L_B \} = 0
\en
where $L_A$ denotes multiplication from the left by the element $A$.
The mixed bracket ${[}, \}$ represents an anticommutator for any two
odd operators and a commutator otherwise. The super Jordan triple
product is defined as \cite{BG,MG80,MG91}:
\eq
(ABC) = A \cdot (B\cdot C) - (-1)^{d_A d_B} B \cdot (A\cdot C)
+ (A\cdot B) \cdot C
\en
The superconformal algebras $g$
 of Jordan superalgebras can be constructed
in complete analogy with the TKK construction for Jordan algebras
 \cite{BG,MG80,MG90}
. One defines a three-graded Lie superalgebra
\eq
g = g^{-1} \oplus g^{0} \oplus g^{+1}
\en
where the elements of the grade $\pm 1$ subspaces are labelled by the elements
of the Jordan superalgebra $J$:
\eqn
U_A &\in& g^{+1} \\
&& \nn  \\
U^A &\in& g^{-1} \nn
\enn
Their supercommutators give the generators $S_A^B$
 belonging to the grade zero subspace
\eq
{[} U_A , U^B \} = (-1)^{d_A d_B} S_A^B \, \in g^0
\en
The remaining supercommutation relations of $g$ turn out to be \cite{MG91}:
\eqn
{[} S_A^B , U_C \}& =& U_{(ABC)} \nn \\
&& \nn \\
{[}U^C , S_A^B \} &=& U^{(BAC)}  \\
&& \nn \\
{[} U_A , U_B \} &=& 0 \nn \\
&& \nn \\
{[} U^A , U^B \} &=& 0  \nn
\enn
Below we list the finite dimensional Jordan superalgebras $J$ as
classified by Kac and their conformal ($G$) and reduced structure groups
($H$) \cite{BG,MG75,MG90}.

\footnotesize

\begin{displaymath}
\begin{array}{|c|c|c|}
\hline
&& \\
J & H & G \\
\hline
&& \\
JA(m^{2}+n^{2}/2mn)  &
SU(m/n)\times SU(m/n) & SU(2m/2n) \\
&& \\
JBC(\frac{1}{2}(m^{2}+m)+(2n^{2}-n)/2mn)  &
SU(m/2n) & OSp(4n/2m) \\
&& \\
JD(m/2n)  &
OSp(m/2n) & OSp(m+2/2n) \\
&& \\
JP(n^{2}/n^{2})  &
SU(n/n) & P(2n-1) \\
&& \\
JQ(n^{2}/n^{2})  &
Q(n-1)\times Q(n-1) \times U(1)_{F} & Q(2n-1) \\
&& \\
JD(2/2)_{\alpha}  &
SU(1/2) & D(2,1;\alpha) \\
&& \\
JF(6/4)  &
OSp(2/4) & F(4) \\
&& \\
JK(1/2)  &
SU(1/2) & SU(2/2) \\
&& \\
\hline
\end{array}
\end{displaymath}
\normalsize
where $U(1)_F$ denotes a fermionic $U(1)$ factor generated by a single
odd generator.  The Jordan algebra of type $X$ with m even and n odd elements
 in Kac's notation was denoted
as $JX(m/n)$ above. For different real forms of the Jordan superalgebra one
obtains different real forms of the conformal superalgebras listed.
 In addition to the above list there exists an infinite
family of ``Hamiltonian `` Jordan superalgebras discovered by Kantor \cite{IK3}
which are exceptional \cite{EZ}. The Jordan superalgebras under the triple
product $(ABC)$ satisfy the graded generalizations of the identities
(\ref{eq:JTS}), which can be taken to be defining identities of super
 Jordan triple systems. To our knowledge, a complete classification of super
Jordan triple systems has not yet appeared in the literature. Such a
classification should readily follow from a classification of three graded Lie
superalgebras. Since the above construction of conformal superalgebras involves
only
the triple product it extends directly to super Jordan triple systems as well.

The  oscillator realization of the conformal and structure
algebras of Jordan algebras and JTS's given in the previous section
can be extended to those of Jordan superalgebras and super JTS's, which
will be generically denoted as $J$. Choose a basis $e_A =(e_a,e_i)$ of $J$
where $e_a$ and $e_i$ are the even and odd basis elements, respectively.
The structure constants of the triple product in this basis are given by
\eq
(e_A e_B e_C) = \Sigma_{AC}^{BD} e_D
\en
We introduce a set of $N$ super oscillators $Z_A = (A_a, \alpha_i)$
 that satisfy
\eqn
{[}Z_A , Z^B \}& =& \delta_A^B  \\
&& \nn \\
{[}Z_A, Z_B \}& =& {[}Z^A, Z^B \} =0 \nn \\
&& \nn \\
A,B,..&=& 1,2,...,N \nn
\enn

 The first $D$ components ($Z_a = A_a, a=1,2,..D$)
 of $Z_A$ are bosonic oscillators
and the remaining $(N-D)$ components $(Z_{(D+i)}= \alpha_i , i=1,2,..N-D)$
 are fermionic.
Again the mixed bracket is an anti-commutator for any two fermionic oscillators
and a commutator otherwise. The following bilinears of the super oscillators
\eq
S_A^B = - \Sigma_{AC}^{BD} Z^C Z_D
\en
close under  supercommutation and form the structure superalgebra of $J$:
\eqn
{[} S_A^B , S_C^D \}& \equiv & S_A^B S_C^D - (-1)^{(d_A+d_B)(d_C+d_D)} S_C^D
S_A^B \\
\nn \\
&& = - \Sigma_{AE}^{BD} S_C^E +(-1)^{(d_A+d_B)(d_C+d_D)} \Sigma_{AC}^{BE}
S_E^D \nn
\enn

The trilinear operators $\Sigma_{CD}^{AB} Z^C Z^D Z_B $ and the linear
operators $Z_A$ together with the bilinears $S_A^B$ generate the superconformal
algebras of the corresponding Jordan superalgebra or the super JTS.
 By realizing the super oscillators as
\eqn
A^a& =& x^a \hspace{1.0cm} A_a = \frac{\partial}{\partial x^a} \\
&& \nn \\
\alpha^i &=& \theta^i \hspace{1.0cm} \alpha_i = \frac{\partial}{\partial
\theta^i}
\enn
where $\theta^i$ are Grassmann coordinates, one obtains a differential
operator realization of the generators of the superconformal algebra. Its
action on the superspace with coordinates $(x^a,\theta^i)$ is equivalent to
the action of the superconformal algebra on the elements $z$ of the Jordan
superalgebra parametrized  as
\eq
z= e_a x^a + e_i \theta^i
\en

\end{document}